\documentclass[10pt,a4paper]{article}

\usepackage{authblk}
\usepackage{amssymb}
\usepackage{hyperref}
\usepackage{graphicx}
\usepackage{array}
\usepackage{siunitx}
\usepackage[usenames, dvipsnames]{color}
\usepackage{color}
\usepackage{subfig}
\usepackage{epsfig}
\usepackage{longtable}
\usepackage[utf8]{luainputenc}

\usepackage{epsfig}
\usepackage{epstopdf}
\usepackage{xspace}
\usepackage{multicol}

\usepackage{makecell}
\usepackage{cite}

\newcommand{\Smilei}{{\sc Smilei}\xspace}

\usepackage[margin=1.9cm]{geometry}

\usepackage{array}
\newcolumntype{L}{>{\centering\arraybackslash}m{3cm}}
\usepackage{multirow}

\hypersetup{%
   colorlinks=true,
   urlbordercolor={1 1 1},
   linkcolor={black},
   raiselinks=false,
citecolor=blue,
urlcolor=blue
}

\newcommand\blfootnote[1]{%
  \begingroup
  \renewcommand\thefootnote{}\footnote{#1}%
  \addtocounter{footnote}{-1}%
  \endgroup
}

\begin{document}

\title{\textbf{Efficient start-to-end 3D envelope modeling for two-stage laser wakefield acceleration experiments}}

\author{F. Massimo$^1$*, A. Beck$^1$, J. Derouillat$^2$,  M. Grech$^3$, M. Lobet$^2$, F. Pérez$^3$, I. Zemzemi$^1$ and A. Specka$^{1}$}
\date{\small{$^1$ Laboratoire Leprince-Ringuet – École polytechnique, CNRS-IN2P3, Palaiseau 91128, France\\
$^2$ Maison de la Simulation, CEA, CNRS, Université Paris-Sud, UVSQ, Université Paris-Saclay, F-91191 Gif-sur-Yvette, France\\
$^3$ Laboratoire d’Utilisation des Lasers Intenses, CNRS, École polytechnique, CEA, Université Paris-Saclay, UPMC Université Paris 06: Sorbonne Universités, F-91128 Palaiseau Cedex, France }}

\maketitle

\blfootnote{*Corresponding author. E-mail address: \href{mailto:massimo@llr.in2p3.fr}{massimo@llr.in2p3.fr}}

\begin{abstract}
Three dimensional Particle in Cell simulations of Laser Wakefield Acceleration require a considerable amount of resources but are necessary to have realistic predictions and to design future experiments. The planned experiments for the Apollon laser also include two stages of plasma acceleration, for a total plasma length of the order of tens of millimeters or centimeters. In this context, where traditional 3D numerical simulations would be unfeasible, we present the results of the application of a recently proposed envelope method, to describe the laser pulse ant its interaction with the plasma without the need to resolve its high frequency oscillations. The implementation of this model in the code \Smilei is described, as well as the results of benchmark simulations against standard laser simulations and applications for the design of two stage Apollon experiments.
\end{abstract}

\section{Introduction}
The maximum electric field sustainable by an accelerating cavity determines the minimum size of a particle accelerator. The breakdown limits of the metallic accelerating cavities in conventional accelerators motivated the accelerator community to find alternative technologies to achieve higher accelerating gradients and thus smaller particle accelerators. One of the most promising of these technologies to accelerate electrons is the Laser Wakefield Acceleration (LWFA), i.e. their acceleration by plasma waves generated in the wake of an intense laser pulse propagating in an under dense plasma \cite{TajimaDawson79,Malka2002,Esarey2009,Malka2012}. The future realization of the Apollon laser \cite{Cros2014}, will pave the way to novel LWFA experiments with high laser power.

The importance of modeling in the LWFA field can hardly be overestimated, since it represents a tool of experimental design, physical prediction, analysis and understanding of the involved phenomena. Nonetheless, LWFA modeling implies sensitive numerical challenges, which require the use of High Performance Computing (HPC) techniques. The state-of-the-art type of codes used to model LWFA is the Particle in Cell (PIC) \cite{BirdsallLangdon2004} code, which samples the plasma distribution function through macro-particles (MP), pushed by and generating electromagnetic (EM) fields defined on a mesh. Vlasov's equation is thus solved following its characteristics, i.e. the MP equations of motion. The PIC method self-consistently updates at each iteration the MP positions and momenta and the EM fields they generate, as well as external EM fields such as a laser pulse.
In LWFA simulations, the largest scales to consider are defined by the size of the plasma accelerator.
For centimeters long accelerating stages, the total propagation time is of the order of hundreds of picoseconds.
On the other hand, the smallest scales are usually defined by the laser.
Indeed, the standard PIC method imposes to resolve the laser wavelength $\lambda_0$ and period $2\pi/\omega_0$ which are respectively of the order of the micron and the femtoseconds.
The discrepancy in scales quickly demonstrates that realistic 3D or quasi-3D \cite{Davoine2008} simulations of this nature are very costly.
They even are beyond reach of standard laser techniques for accelerating stages reaching tens of centimeters as is planned for multi-stage LWFA. Henceforth, we will call  standard laser techniques, standard laser simulations those not using physical approximations to speed-up the calculations. Examples of these techniques include but are not limited to quasi-static approximation \cite{Mora1997}, Azimuthal Fourier decomposition \cite{Lifschitz2009}, boosted frame techniques \cite{Vay2007} or hybrid models \cite{Benedetti2010,Benedetti2012,MassimoJCP2016}.

A possible technique to reduce the computing cost of LWFA simulations consists in using a description of the laser pulse that takes into account only its complex envelope, with length and transverse size of the order of the plasma wavelength $\lambda_p \gg \lambda_0$, without the need to resolve each optical cycle of the laser in time and space \cite{Mora1997,Gordon2000,Huang2006,Cowan2011,Benedetti2012,Helm2016}. The code WAKE was one of the first codes with a laser envelope model to simulate laser wakefield scenarios \cite{Mora1997}. However, the quasi-static approximation (QSA) of the code prevented the simulation of electrons injection. Time explicit (i.e. without QSA) envelope codes were developed,  with implicit schemes to solve the evolution equation of the laser envelope \cite{Gordon2000,Cowan2011,Benedetti2012,Helm2016}. The use of implicit solvers to solve an envelope equation requires non-trivial parallelization techniques to be used in 3D, as they often require the inversion of a matrix representing a transverse differential operator acting on the envelope \cite{Cowan2011,Benedetti2012,Helm2016,Benedetti2018}. Recently, a time explicit 3D envelope code with an easily parallelizable explicit solver for the envelope equation has been developed in the PIC code ALaDyn \cite{Terzani2019}.

In this paper, we present applications of this technique now also implemented in the PIC code \Smilei \cite{Smilei2018,Beck2019}, to demonstrate its suitability for studies oriented to the LWFA experiments planned for Apollon.

In the second section, we briefly review the envelope model's approximations and equations. In the third section a numerical simulation of a second stage experiment is discussed, comparing its results with a standard laser simulation. In the fourth section, five simulations of possible working points for the first injecting stage for Apollon are presented. In the Appendix, the envelope model equations and their solution in \Smilei are summarized. 

\section{Review of the envelope model}
\begin{figure}[htbp]
\begin{center}
\includegraphics[scale=0.4]{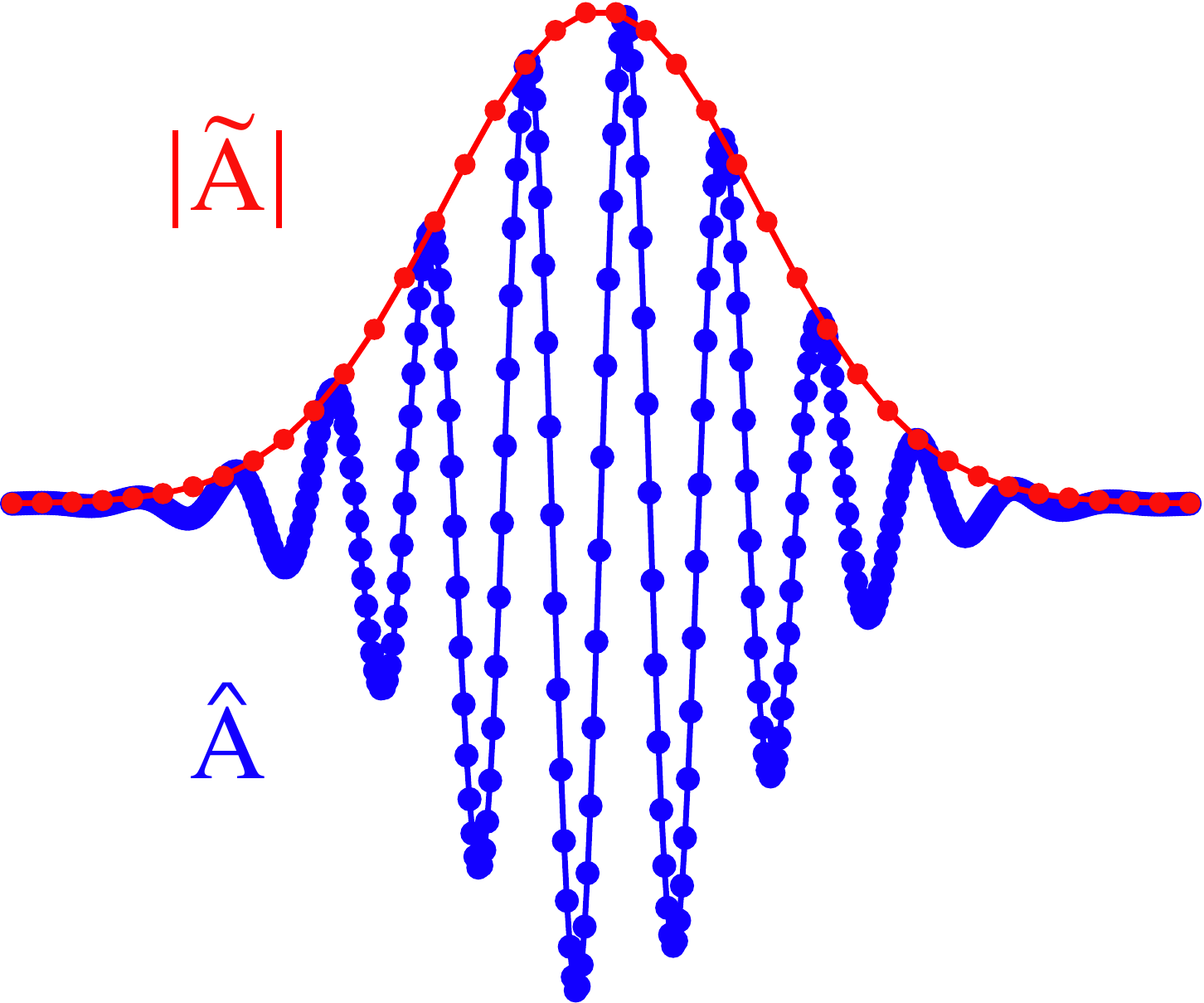}
\caption{Blue line: laser vector potential component $\hat{A}$ along the polarization direction. Red line: the module of its complex envelope $|\tilde{A}|$. The envelope is sampled by a number of points smaller by a factor ten compared to $\hat{A}$.}
\label{fig:envelope_concept}
\end{center}
\end{figure}

In many typical situations for LWFA, the spatial and temporal scales of interest (e.g. the wavelength of the accelerating plasma wave, scaling as $\lambda_p$) are significantly larger than the scales related to the laser central wavelength $\lambda_0$. In these cases, if the laser pulse is much longer than $\lambda_0$, the computation time can be substantially reduced if one may sample only the laser envelope characteristic length (which normally scales as $\lambda_p$ as well in typical LWFA setups) instead of $\lambda_0$, as depicted in Fig. \ref{fig:envelope_concept}. With this lower resolution, the region of interest for the electron acceleration, i.e. the end of the plasma ``bubble" in the first period of the plasma wave behind the laser, would still be correctly described in most of the situations of interest for LWFA. Indeed, the bubble formation is triggered by an averaged interaction of the particles with the laser, i.e. the ponderomotive force \cite{Quesnel1998}, which derives from the laser envelope.

The envelope model implemented in \Smilei is similar to the one first demonstrated in the PIC code ALaDyn, including the same solver for the envelope equation in laboratory frame coordinates and the ponderomotive solver for the particles' equations of motion presented in \cite{Cowan2011,Terzani2019}.
In the following, the equations of this model are presented. Various numerical schemes for their solution are detailed in \cite{Terzani2019} and those implemented in \Smilei are reviewed in \ref{appendix}. 
Henceforth, normalized units will be used for all quantities (choosing $k_0^{-1}$ as the normalized unit length, $c$ as the normalized unit velocity, etc.).

The fundamental assumption of the model is the description of the laser pulse vector potential in the complex polarization direction $\hat{A}(\mathbf{x},t)$ as a slowly varying envelope $\tilde{A}(\mathbf{x},t)$ modulated by fast oscillations at wavelength $\lambda_0$:

\begin{equation}\label{envelope_form}
\hat{A}(\mathbf{x},t)=\textrm{Re}\left[\tilde{A}(\mathbf{x},t)e^{i(x-t)}\right],
\end{equation}

Thus, in general any physical quantity $A$ will be therefore given by the summation of a slowly varying part $\bar{A}$ and a fast oscillating part $\hat{A}$, i.e. $A=\bar{A}+\hat{A}$, where $\hat{A}$ has the same structure as in Eq. \ref{envelope_form} (we are using the same notation used in \cite{Cowan2011}). The laser vector potential in vacuum would have only the fast oscillating part ($\bar{A}=0$).
In this context, ``slowly varying" means that the space-temporal variations of $\bar{A}$ and of the envelope $\tilde{A}$ of the fast oscillating part are small enough to be treated perturbatively with respect to the ratio $\epsilon=\lambda_0/\lambda_p$, as described in detail in \cite{Mora1997, Quesnel1998, Cowan2011}. The laser envelope transverse size $R$ and longitudinal size $L$ are thus assumed to scale as $R \approx L \approx \lambda_0 / \epsilon$ \cite{Mora1997, Quesnel1998}.
As described thoroughly in the same references, the action of the laser envelope on the plasma particles is modeled through the addition of a ponderomotive force term in the particles equations of motion. This term, not representing a real force, is a term rising from an averaging process in the perturbative treatment of the particles motion over the laser optical cycles. The particles equations of motions (in this case for electrons) thus read:
\begin{eqnarray}
\frac{d\mathbf{\bar{x}}}{dt}&=&\frac{\bar{\mathbf{p}}}{\bar{\gamma}} \\
\frac{d\mathbf{\bar{p}}}{dt}&=&-\left(\mathbf{\bar{E}}+\frac{\bar{\mathbf{p}}}{\bar{\gamma}}\times\mathbf{\bar{B}}\right)-\frac{1}{\bar{\gamma}}\nabla\Phi \\
\bar{\gamma}&=&\sqrt{1+|\mathbf{\bar{p}}|^2+\Phi}
\end{eqnarray}
where $\mathbf{\bar{x}}$, $\mathbf{\bar{p}}$, $\mathbf{\bar{E}}$, $\mathbf{\bar{B}}$ are the slow-varying parts of the particles positions and momenta and of the electric and magnetic field. The ponderomotive potential is defined as $\Phi=|A|^2/2$. The Lorentz factor in the usual motion equations is replaced by the ponderomotive Lorentz factor $\bar{\gamma}$ \cite{Mora1997, Quesnel1998}.
The quantities stored in an envelope simulation in \Smilei are the slowly varying MP positions and momenta $\mathbf{\bar{x}}$, $\mathbf{\bar{p}}$, the slowly varying electromagnetic fields $\mathbf{\bar{E}}$, $\mathbf{\bar{B}}$, the envelope $\tilde{A}$, the ponderomotive potential $\Phi$ and its gradient $\nabla\Phi$. In a standard laser simulation, the complete (high frequency part and low frequency part) MP positions and momenta $\mathbf{x}$, $\mathbf{p}$ and electromagnetic fields $\mathbf{E}$, $\mathbf{B}$ are stored. 

The evolution of the laser pulse envelope is derived combining d'Alembert's inhomogeneous equation and Eq. \ref{envelope_form}:

\begin{equation}\label{envelope_equation}
\nabla^2 \tilde{A}+2i\left(\partial_x \tilde{A} + \partial_t \tilde{A}\right)-\partial^2_t\tilde{A}=\chi \tilde{A}.
\end{equation}

The function $\chi$ represents the plasma susceptibility, which in a PIC code can be computed similarly to the charge density \cite{Benedetti2012,Terzani2019}. This term takes into account the effect of the plasma on the laser propagation, and is necessary to model phenomena of self-focusing \cite{Sun1987}. The derivation of Eq. \ref{envelope_equation} assumes that the high frequency contribution of the scalar potential can be ignored \cite{Esarey2009,Cowan2011}.

In \Smilei, no further assumption on the envelope is made and the full form of Eq. \ref{envelope_equation} is solved.  Thus, Eq. \ref{envelope_equation}, solved in ALaDyn and \Smilei, is physically equivalent to the envelope equation in the code INF\&RNO (Eq. 1 in \cite{Benedetti2012,Benedetti2018}), where it is written in comoving coordinates and solved with an implicit  numerical scheme detailed in \cite{Benedetti2018}. As explained in \cite{Benedetti2012}, we remark that retaining all the terms in the derivation of the envelope equation allows in principle to use this model in conjunction with the Lorentz boosted frame method \cite{Vay2007}, since the form of Eq. \ref{envelope_equation} is invariant under Lorentz transformations.

The slowly varying electromagnetic fields $\bar{\mathbf{E}}$, $\bar{\mathbf{B}}$ are updated solving Maxwell's equations, with the slowly varying current densities $\bar{\mathbf{J}}$ as source terms. Since the form of Maxwell's equations is unaltered, a Finite Difference Time Domain (FDTD) scheme \cite{Yee1966} is used to evolve these fields in the simulations described hereafter. The deposition of $\bar{\mathbf{J}}$ is performed with the charge conserving scheme by Esirkepov \cite{esirkepov:CPC2001}.

\section{Benchmark case study: second stage simulations}\label{second_stage}
An envisioned experiment for Apollon is the injection of electrons from a first nonlinear plasma stage and their transport line towards a second plasma stage, where they will be accelerated by plasma waves in the weakly nonlinear regime. The lower accelerating gradients of weakly nonlinear regimes imply the requirement of a large accelerating length for the second plasma stage to have a sensitive energy gain. This represents a considerable challenge from the experimental point of view, but also for the numerical simulation, which would be unfeasible with a standard laser simulation techniques.
In order to demonstrate the suitability of the envelope model for second stage simulations, we report a comparison between a standard laser and an envelope simulation. The physical setup is given by a laser pulse injected in a plasma and an idealized Gaussian electron beam injected from outside the plasma, as for an external injection experiment into a second stage. The ideal laser, plasma and beam parameters are the same as in \cite{XiangKunLi2018} and are briefly recalled in the following.

The Gaussian laser pulse, linearly polarized in the $y$ direction, has a waist size $w_0=45$ $\mu$m. Its Gaussian temporal profile has an initial FWHM duration $\tau_0=108$ fs in intensity and peak normalized field amplitude $a_0=\sqrt{2}$. The idealized plasma profile is given by a transversely parabolic profile $n_e(r)=n_0\left(1+\frac{\Delta n}{n_0}\frac{r^2}{r^2_0}\right)$ (with $r$ the distance to axis, $n_0=1.5\cdot10^{17}$ cm$^{-3}$, $\frac{\Delta n}{n_0}=0.25$, $r_0=45$ $\mu$m), starting from the initial right boundary of the moving window, which has longitudinal and transverse dimensions $L_x=1400$ $k_0^{-1}$ and $L_y=L_z=1600$ $k_0^{-1}$. The laser is focused at the beginning of the plasma profile, and its initial center is chosen at a distance $2\sqrt{2}\thinspace\tau_0$ from the plasma. The longitudinal cell size for the standard laser simulation is $\Delta x_{\rm laser} = 0.196$ $k_0^{-1}$ and the transverse cell size is $\Delta y = \Delta z = 3$ $k_0^{-1}$. The time step has been chosen as $\Delta t_{\rm laser} = 0.95 \thinspace\Delta x_{\rm laser}$. For the envelope simulation, the longitudinal grid cell size and the integration time step have been set to $\Delta x_{\rm envelope}=16\thinspace\Delta x_{\rm laser}$, $\Delta t_{\rm envelope}=0.8\Delta x_{\rm envelope}$ respectively.
In both the simulations, 8 particles per cell have been used.

According to our tests, the larger longitudinal cell size and integration timestep in the envelope simulation seed the numerical Cherenkov radiation (NCR) more quickly than in the standard laser simulations, as expected from the underlying theory \cite{Lehe2013}. This numerical artifact has detrimental effects on the beam emittance after long propagation distances. To partially cope with the NCR, we use a binomial filter (2 passes) on the current densities at each iteration \cite{Vay2015}. An advantage of the envelope model, as well as of boosted frame simulations as explained in \cite{Vay2015}, is that the laser is not modeled with high frequency oscillations, so low-pass filtering does not risk to damp the high frequency phenomena close to the laser as it would do in a standard laser simulation \cite{Vay2015}. Therefore, in the standard laser simulation no low-pass filter is used.

The electron beam injected after the laser has an ideal Gaussian density profile in all the phase space sub-planes and a total charge of $Q=30$ pC. Its mean energy is $150$ MeV, with a $0.5\%$ rms energy spread. The beam transverse rms size is $1.3$ $\mu$m, its longitudinal rms size $2$ $\mu$m, with a transverse normalized emittance of $1$ mm-mrad. The beam initial position is at a distance $0.75\lambda_p$ after the laser pulse, in a phase with both a focusing and an accelerating field in the wake of the laser in the plasma. All the electron beam MP carry the same electric charge, given by $Q/N_p$, where $N_p=10^6$ is the number of MP which sample the beam. To self-consistently initialize the EM fields of an electron beam that is already relativistic at the beginning of the simulation, the procedure described in \cite{Vay-PoP2008,Massimo2016} was implemented in \Smilei.
To summarize, at the beginning of the simulation the ``relativistic Poisson's equation" is solved, once the initial beam charge density $\bar{\rho}$ is computed:
\begin{equation}
\left( \frac{1}{\gamma^2_0}\partial^2_x+\nabla_{\perp}^2\right) \bar{\Phi} = -\bar{\rho}.
\end{equation}
Then, the low frequency EM fields $\mathbf{\bar{E}}$, $\mathbf{\bar{B}}$ at the same time step $t=0$ are computed:
\begin{eqnarray}
\mathbf{\bar{E}}&=&\left( -\frac{1}{\gamma_0^2}\partial_x , -\partial_y, -\partial_z  \right)\bar{\Phi},\\ \mathbf{\bar{B}}&=&\beta_0\mathbf{\hat{x}}\times\mathbf{\bar{E}}.
\end{eqnarray}
The quantities $\gamma_0=1/\sqrt{1-\beta^2_0}$ and $\beta_0$ represent the initial mean beam Lorentz factor and initial normalized speed respectively. The quantities $\bar{\rho}$ and $\bar{\Phi}$ are centered on the primal grid in all the $x$, $y$, $z$ directions. The derivatives in the previous equations are computed through finite differences. In order to provide the properly space-centered fields for the FDTD scheme, the magnetic field is then spatially interpolated in the Yee cell. Besides, in order to provide the properly time-centered initial conditions for the FDTD scheme, the magnetic field at time $-\Delta t/2$ is found using a backward FDTD ``advance" by $-\Delta t/2$.

Table \ref{table:beam_params_2nd_stage} reports the electron beam parameters obtained by the two simulations after a propagation distance of $15$ mm. We note a very good agreement for certain parameters as the mean energy, the transported charge and the beam duration after this distance. The most striking differences are found in the energy spread and in the transverse plane parameters as the rms sizes $\sigma_i$ ($i=y,\thinspace z$) and normalized emittances $\varepsilon_{n,i}$. The normalized emittance is defined as $\varepsilon_{n,i}=\sqrt{\sigma_i^2\sigma_{p_i}^2-\sigma_{ip_i}^2}$ ($i=y,\thinspace z$), where $\sigma_{p_i}$ and  $\sigma_{ip_i}$ are the beam rms momentum spread and covariance between the coordinate $i$ and the momentum $p_i$ in the transverse planes $y-p_y$ and $z-p_z$.

The differences in the beam emittance, which is conserved along the propagation in the envelope simulation, is mainly due to the mitigation of NCR thanks to the smoothing operated on the current density. The possibility of low-pass filtering, and thus reduction of the effects of this numerical artifact plaguing standard LWFA PIC simulations, without compromising the laser propagation represents an important advantage of envelope simulations.
PIC simulations of LWFA not using envelope need more advanced techniques to cope with NCR, like {\it ad-hoc} solvers for Maxwell's Equations \cite{Karkkainen2006,Lehe2013,Cowan2013} instead of the FDTD solver, or pseudo-spectral electromagnetic solvers \cite{Jalas2017} coupled to the Galilean frame method \cite{Lehe2016}. 

\begin{figure}[htbp]
\begin{center}
\includegraphics[scale=0.5]{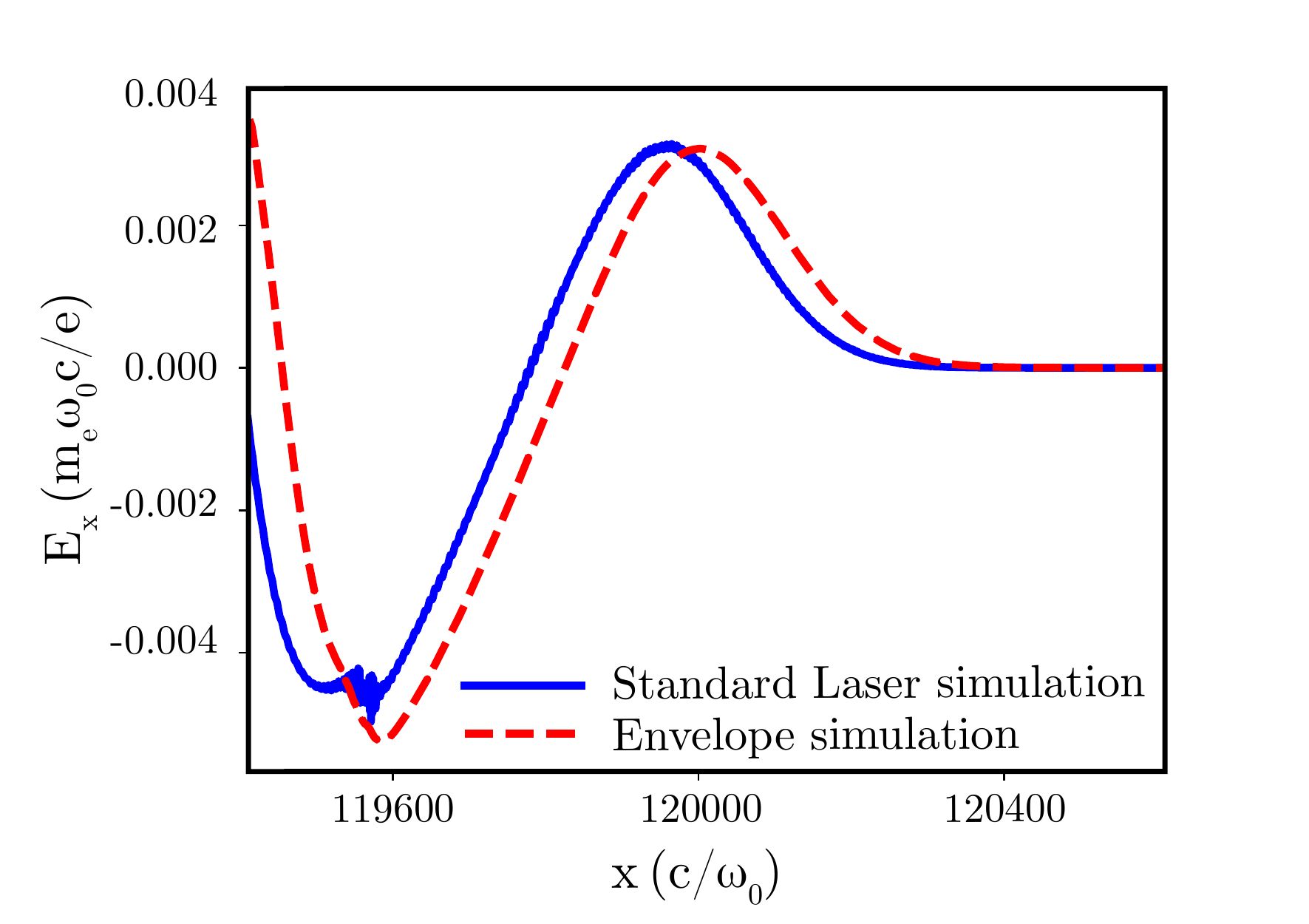}
\caption{Longitudinal electric field $E_x$ on axis after $15$ mm of propagation, comparing the results obtained with the envelope model and the results obtained through a standard laser simulation. }
\label{fig:Ex1D_2nd_stage}
\end{center}
\end{figure}

\begin{figure}[htbp]
\begin{center}
\includegraphics[scale=0.5]{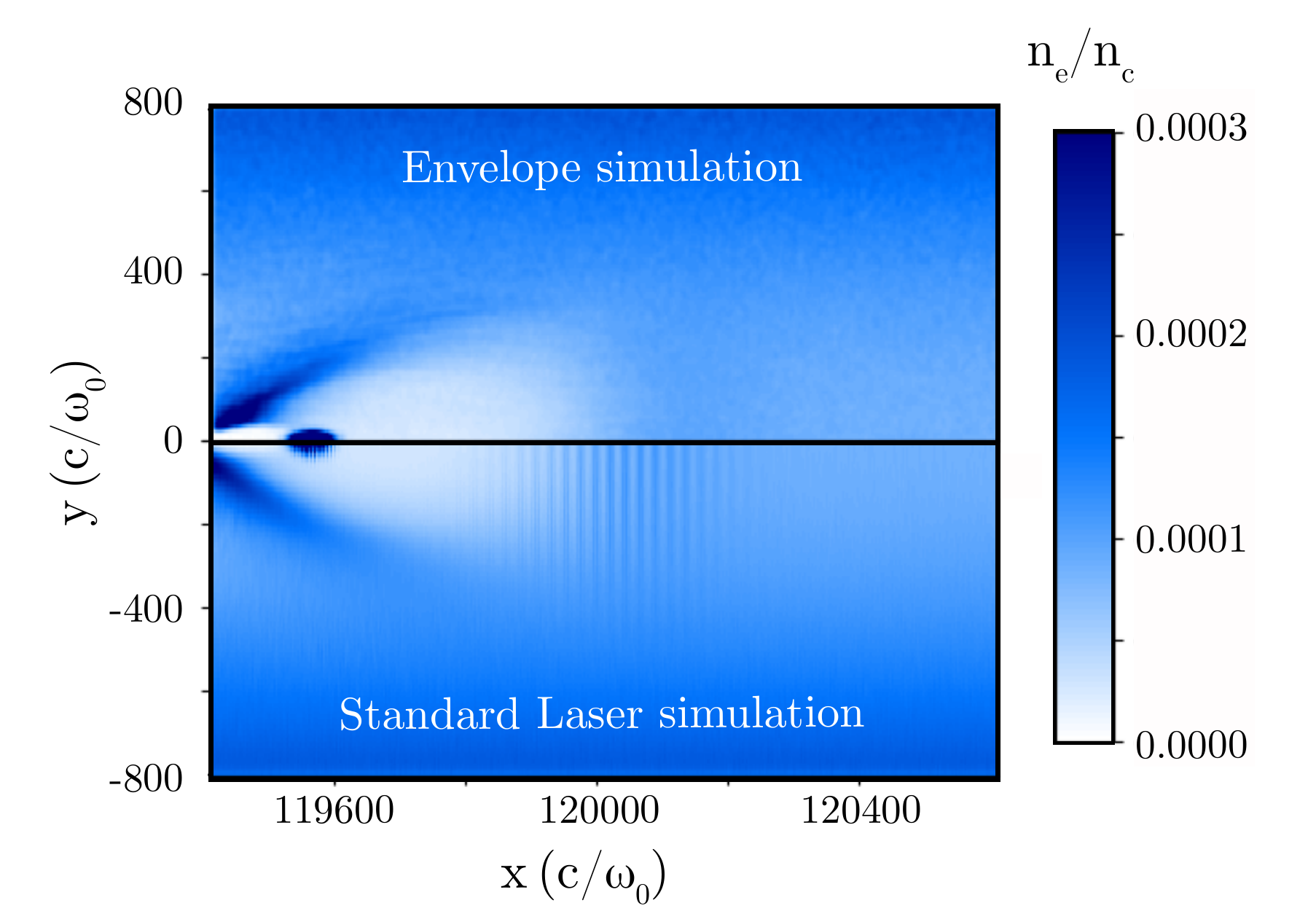}
\caption{Plasma density after $15$ mm of propagation. Top panel: results obtained with the envelope model. Bottom panel: results obtained through a standard laser simulation.}
\label{fig:Rho2D_2nd_stage}
\end{center}
\end{figure}

\begin{figure}[htbp]
\begin{center}
\includegraphics[scale=0.5]{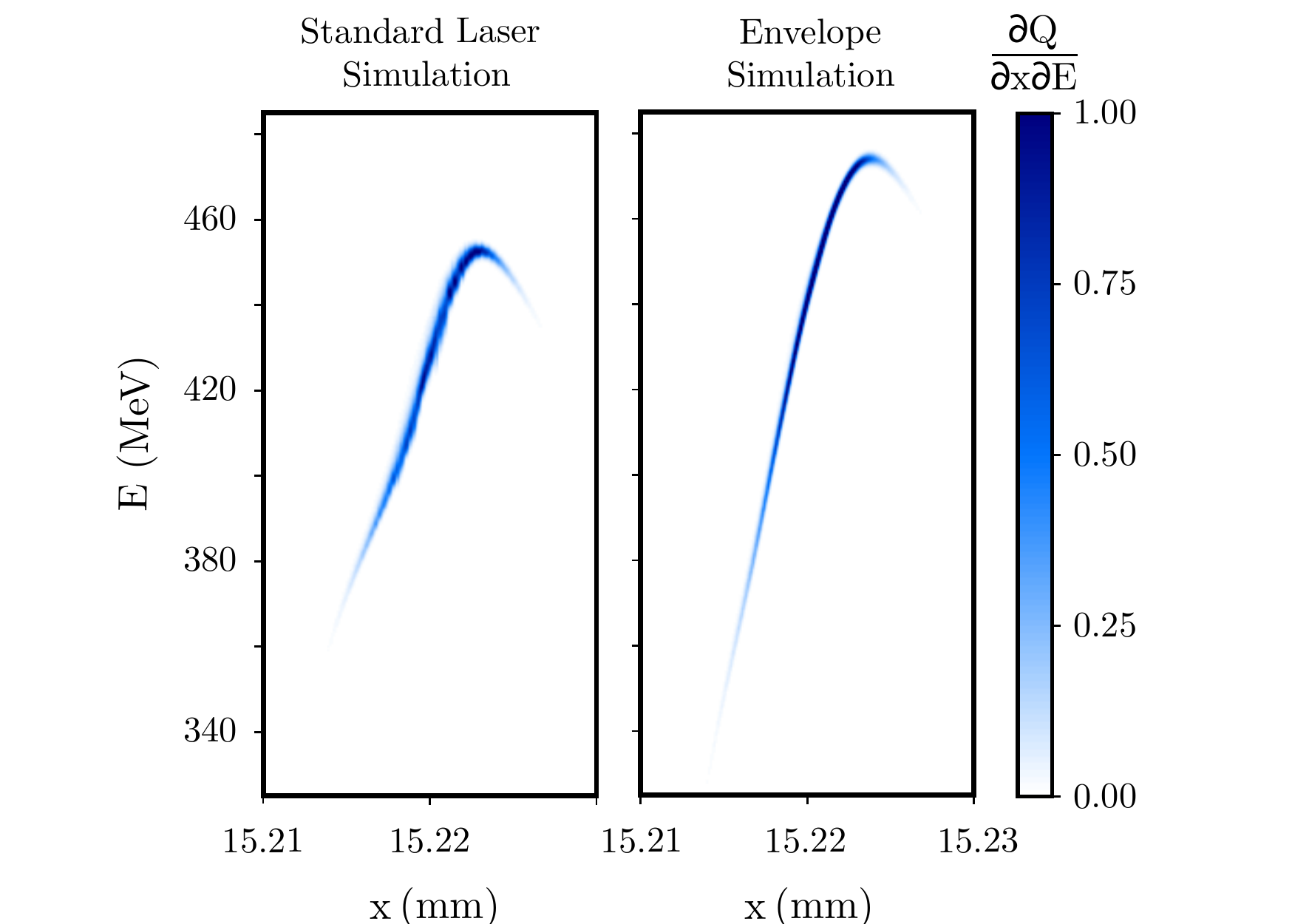}
\caption{Normalized longitudinal phase space distribution of the externally injected and LWFA-accelerated electron beam after a propagation distance of $15$ mm.}
\label{fig:spectrum_2nd_stage}
\end{center}
\end{figure}

The differences in energy can be explained noting that the envelope solver has a minor numerical dispersion compared to a standard FDTD scheme \cite{Terzani2019}. It was already shown in \cite{Cowan2013} that differences in the dispersion relation in the laser/envelope propagation scheme can lead to significant differences in the longitudinal phase space evolution of laser wakefield-accelerated beams. 
Indeed, as explained in \cite{Cowan2013}, less accurate description of the laser propagation speed because of numerical dispersion can lead to an overestimation of the dephasing effect hence an underestimation of the accelerating field. From Fig. \ref{fig:Ex1D_2nd_stage}, where the longitudinal electric field on axis is compared at $15$ mm, it is evident that the laser pulse is moving faster in the envelope simulation, and after a long simulation time the difference in the traveled distance becomes sensitive. Instead, the electron beam moves at the same velocity (close to $c$) in both simulations and is found at the same position (see Fig. \ref{fig:Rho2D_2nd_stage}) at the same simulation time. Thus, the electron beam is subject to a numerically exaggerated dephasing in the standard laser simulation. As can be seen from  Fig. \ref{fig:Ex1D_2nd_stage}, the beam loading is sensitive to the different wakefield phases where the electron beam is found in the two simulations, where the laser propagates ad different speed. These differences in beam loading and in dephasing yield a lower energy and a lower energy spread in the standard laser simulation, clearly visible in the longitudinal phase space distribution (Fig. \ref{fig:spectrum_2nd_stage}).

To summarize, in these particular conditions the simulation of the second stage can greatly benefit from a modeling based on the envelope approximation. Compared to standard laser simulations, calculations using laser envelopes can use digital filters to cope with NCR 
without damping the relevant physics near the laser and the reduced numerical dispersion can yield more accurate results. Besides, the envelope simulations only need a fraction ( 5\% in this case) of the computing resources required by a standard laser simulation. 
The necessary computing resources for this simulation up to 15 mm with a standard laser are not easily available normally and were provided by the Grand Challenge "Irene" 2018 project of GENCI.

\begin{table}[hbtp]
\centering
\renewcommand{\arraystretch}{1}
\begin{tabular}{l|c|c|c|}
\cline{2-4}
& \multicolumn{1}{l|}{Initial values} & \multicolumn{1}{l|}{Standard Laser (15 mm)} & \multicolumn{1}{l|}{Envelope (15 mm)} \\ \hline
\multicolumn{1}{|l|}{$Q$($pC$)} & 30 & 29.98 & 29.94 \\ \hline
\multicolumn{1}{|l|}{$\sigma_x$ ($\mu$m)} & 2.0 & 2.0 & 2.0 \\ \hline
\multicolumn{1}{|l|}{$\sigma_y$ ($\mu$m)} & 1.3 & 1.5 & 1.0 \\ \hline
\multicolumn{1}{|l|}{$\sigma_z$ ($\mu$m)} & 1.3 & 1.4 & 1.0 \\ \hline
\multicolumn{1}{|l|}{$\varepsilon_{n,y}$ (mm-mrad)} & 1.0 & 2.0 & 1.0 \\ \hline
\multicolumn{1}{|l|}{$\varepsilon_{n,z}$ (mm-mrad)} & 1.0 & 2.1 & 1.0 \\ \hline
\multicolumn{1}{|l|}{$E$ (MeV)} & 150 & 427 & 438 \\ \hline
\multicolumn{1}{|l|}{$\sigma_{E}/E$ (\%)} & 0.5 & 4.7 & 6.4 \\ \hline
\end{tabular}
\caption{Beam parameters: charge $Q$, rms sizes $\sigma_i$ ($i=x,y,z$), normalized emittances $\varepsilon_{n,i}$ ($i=y,z$), mean energy $E$, rms energy spread $\sigma_{E}/E$. First column: beam parameters at the beginning of the simulation. Second and third columns: beam parameters after 15 mm of propagation in the standard laser simulation and envelope simulation.}
\label{table:beam_params_2nd_stage}
\end{table}


\section{Single stage simulations for Apollon}\label{first_stage}

This Section illustrates the interest of the envelope model to run simulation campaigns applied to real experimental setups.

In Section \ref{second_stage}, we have compared the results of a standard laser simulation and an envelope simulation for a long distance LWFA benchmark in a weakly nonlinear regime. Examples of the use of time explicit envelope models for highly nonlinear regimes with self-injection, like those discussed in this Section, can be found for example in \cite{Benedetti2010,Cowan2011}.

\begin{figure}[htbp]
\begin{center}
\includegraphics[width=0.6\textwidth]{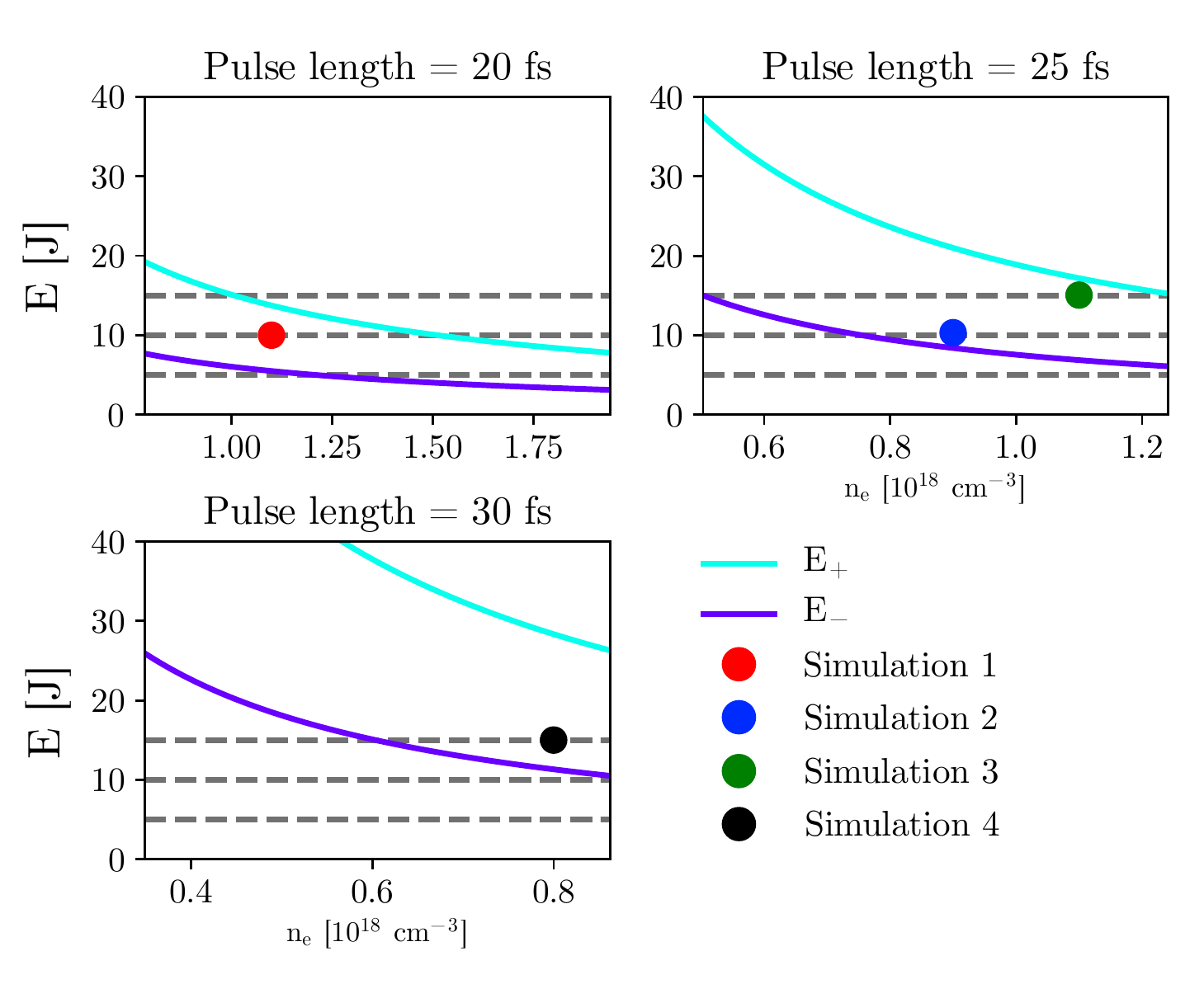}
\caption{Density window prescribed by pulse length.
 Minimum energy $E_{-}$ and maximum energy $E_{+}$ in order to guarantee proper self-focusing and propagation are plotted as functions of the plasma density.
 The plasma density ranges between $n_{-}$, in order to have self-focusing effectively counter-balancing diffraction, and  $n_{+}$ above which laser depletion makes the acceleration inefficient and unstable.
 Optimum parameters are located at the center of the delimited zone.
 Accessible parameters with energies of 5, 10 and 15 J are plotted as gray dashed lines.
 Colored dots mark the parameters chosen for the scan simulations.}
\label{fig:parameter_choice}
\end{center}
\end{figure}

A set of four 3D simulations is performed in order to probe the parameter space accessible to the first Apollon shots. 
The very first experiments will be done with increasing laser energy starting from 5 J and up to 15 J.
We expect the laser pulse duration to be between 20 and 30 fs.
In order to fix parameters for a simple single stage LWFA experiment within this window of laser pulse energy and length, one can use the criteria defined in \cite{beck2014}.
The plasma density window is limited by the diffraction-limited propagation on one side $\omega_p \tau_L \ge 1$, and the depletion-limited propagation on the other $\omega_p \tau_L \le \pi/2$.
And once a density is chosen, the minimum power is $P_- = 10 P_{cr}$ in order for the relativistic self-focusing to counter balance diffraction. We remind that $P_{cr}=16.2(n_c/n_0)$ GW, where $n_0$ and $n_c$ are respectively the electron and critical density respectively \cite{Sun1987}. 
The laser wavelength is $\rm \lambda_0 = 0.8\ \mu m$.
The maximum power is $P_+ = 25 P_{cr}$ in order to stay in the self-guided propagation without critical oscillations of the laser spot size.
The laser pulse length being already fixed, the criteria on $P$ translates directly into a condition on the energy.
This is illustrated in Fig. \ref{fig:parameter_choice} which shows, for different pulse lengths, the minimum and maximum energy as a function of density within the prescribed density window.
It is worthwhile to note that, as the pulse becomes shorter in time, the energy window narrows and the optimal energy decreases.
Future high energy LWFA experiments based on a similar self-guided laser pulse willing to reach high charges or energy should therefore make use of longer laser pulses.

The waist size has been chosen as $40$ $\mu$m and focused at the start of the plasma in all the simulations in order to better compare the various working points. The plasma density of all the simulations was defined with a $300$ $\mu$m upramp from $0$ to the plateau density value $n_0$, reported in Table \ref{tab:scan_parameters}. After the upramp, the plasma density remains constant. Besides, the plasma density is chosen uniform in the transverse direction, to rely only on relativistic self-focusing for the laser guiding \cite{Sun1987}. In all the scan simulations, 8 particles per cell have been used. The integration timestep was always set to $\Delta t=0.9 \thinspace\Delta x$, where $\Delta x$ is the longitudinal mesh cell size. The transverse mesh cell size has been set to $\Delta y = \Delta z = 3.5$ $k_0^{-1}$ for all the simulations. As in the simulations reported in Section \ref{second_stage}, we used a binomial filter on the current densities at each iteration (2 passes). Again, this smoothing aims at limiting the NCR impact on the simulation. 

We remark that a similar set of equivalent 3D simulations would have been unfeasible with standard laser simulations.
For reader's convenience, the parameters of the four simulations have been summarized in table \ref{tab:scan_parameters}, including the longitudinal mesh cell size.

\begin{table}[htbp]
\centering
\renewcommand{\arraystretch}{1}
\begin{tabular}{|c|c|c|c|c|c|c|}
\hline
Simulation & Pulse duration [fs] &  $a_0$ & $n_0$  $[10^{18}$ cm$^{-3}]$ & $P/P_{cr}$  & Laser Energy [J] & $\Delta x$ [$k_0^{-1}$]\\ \hline\hline
       \textcolor{red}{1}          &      20          &      2.96   &   1.1  &  18.3  & 10 & 0.5 \\ \hline
       \textcolor{blue}{2}          &      25          &      2.64   &   0.9  &  12.0  & 10& 0.667 \\ \hline
       \textcolor{OliveGreen}{3}          &      25          &      3.24   &   1.1  &  22.0    & 15 & 0.667   \\ \hline
       4          &      30          &     2.96   &   0.8  &  13.3  & 15 & 0.8  \\ \hline
\end{tabular}
\caption{Laser and Plasma Parameters for the scan simulations. The waist size has been chosen as $40$ $\mu$m for all the four simulations. The laser strength parameter\cite{Esarey2009} is denoted by $a_0$. The longitudinal mesh cell size $\Delta x$ is also reported. The transverse mesh cell size has been set to $\Delta y = \Delta z = 3.5$ $k_0^{-1}$ and the integration timestep to $\Delta t=0.9 \thinspace\Delta x$.}\label{tab:scan_parameters}
\end{table}

\begin{figure}[htbp]
\begin{center}
\includegraphics[scale=0.5]{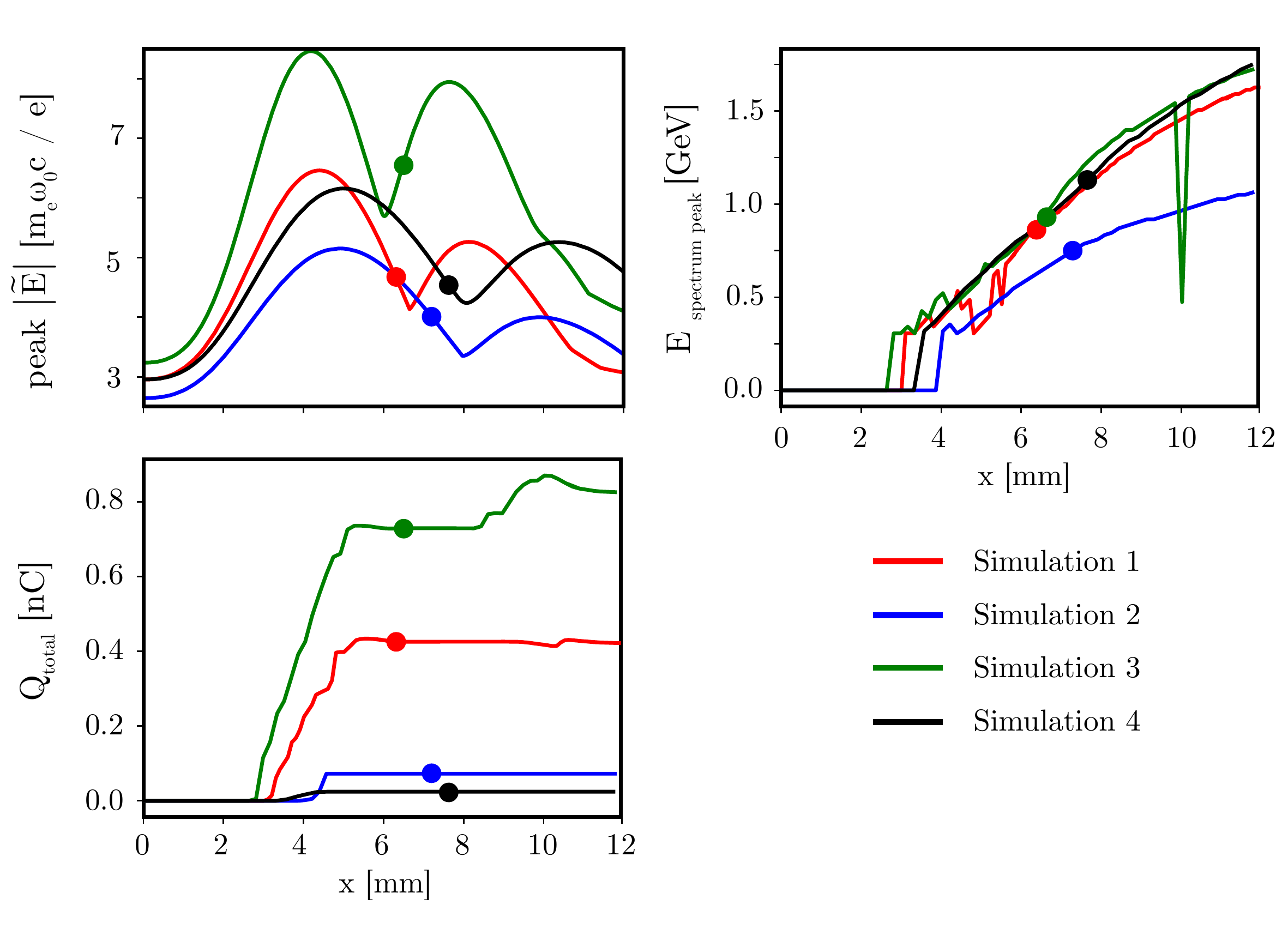}
\caption{
Top left panel: Evolution of the absolute value of the laser peak electric field envelope $|\tilde{E}|$. Top right panel: evolution of the energy of the electron spectrum peak $E_{spectrum\thinspace peak}$ for particles with energy above $300$ MeV.
Bottom left panel: Evolution of the the total charge $Q_{tot}$ above $300$ MeV. 
In all the panels, the colored markers highlight the “optimum" laser propagation distance, before the electron beam starts its rotation in the longitudinal phase space
}
\label{fig:a0_E_Q}
\end{center}
\end{figure}

Figure \ref{fig:a0_E_Q} (top left panel) depicts the evolution of the peak absolute value of the electric field envelope $|\tilde{E}|$ for the four simulations.
From the definition of the envelope $\tilde{A}$ of the vector potential $\hat{A}$ along the polarization direction (Eq. \ref{envelope_form}), this quantity is defined as the amplitude of the envelope of $\hat{E}=-\partial_t \hat{A}$, i.e. $\tilde{E}=-(\partial_t-i)\tilde{A}$.
As desired, the ratios $P/P_{cr}\gg1$ lead to relativistic self-focusing \cite{Sun1987}, which suddenly enlarges the bubble behind the laser, triggering electron injection.
The importance of dark current (low energy tail in electrons energy distribution) depends strongly on the set of parameters used.
An \textit{ad hoc} definition of the beam is used here in order to separate it from the dark current when necessary.

After injection, the energy distribution of the electrons shows a distinct peak at $E_{spectrum\thinspace peak}$ and a full width half maximum $\Delta E$.
We define the injected beam, at a given time, as the electrons having an energy within an interval of width $2\Delta E$ around $E_{spectrum\thinspace peak}$.

Figure \ref{fig:a0_E_Q} also shows the evolution of the total charge $Q_{tot}$ above $300$ MeV (bottom left panel) and the energy of the spectrum peak $E_{spectrum\thinspace peak}$ (top right panel).
As expected, injection occurs when the laser self-focuses \cite{beck2014}.
Since secondary injections can create additional peaks in the electron spectrum, the energy of the peak $E_{spectrum\thinspace peak}$ can vary suddenly when a peak overtakes another one and becomes the new maximum of the distribution (see red line or green line of the top right panel in Figure \ref{fig:a0_E_Q}).


\begin{figure}[htbp]
\begin{center}
\includegraphics[scale=0.65]{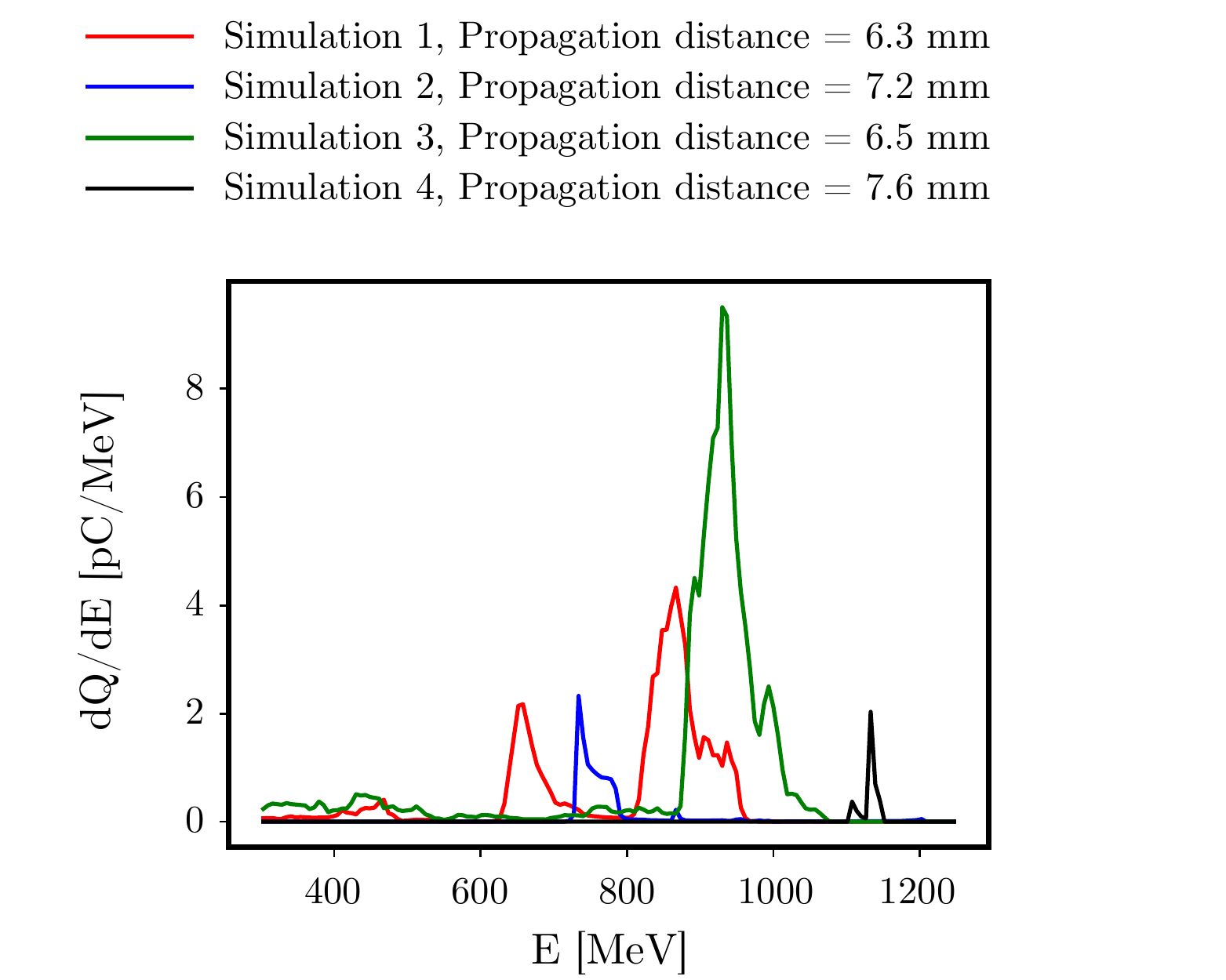}
\caption{Energy spectrum of the single stage simulations at the “optimum" laser propagation distance.
The cut-off energy is 300 MeV. 
}
\label{fig:Energy_distribution}
\end{center}
\end{figure}

To provide interesting working points for the Apollon experiments, in Table \ref{table:beam_params_1st_stage} we report the injected beam parameters after a certain laser propagation distance, different for each simulation. This optimal distance has been chosen by carefully following the evolution of the injected beam in the longitudinal phase space $x-p_x$.
At that point, the electron beam reaches the maximum spectral density.
It corresponds to the moment when phase space rotation minimizes the beam energy spread \cite{Lu2007}.
As seen on Fig. \ref{fig:a0_E_Q}, the acceleration process goes on past this point but at the cost of a reduced beam quality and possible additional dark current.
For all the simulations, this optimal propagation distance is found around 6-7 mm.
In all these working points, the spectrum peak energy is around 1 GeV, with an energy spread lower than $10\%$. 
Fig. \ref{fig:Energy_distribution} reports the electron energy spectrum for all the simulations at this “optimum" laser propagation distance.

\begin{table}[hbtp]
\centering
\renewcommand{\arraystretch}{1}
\begin{tabular}{l|c|c|c|c|}
\cline{1-5} 
\multicolumn{1}{|l|}{Simulation } &
\multicolumn{1}{c|}{\textcolor{red}{1}} & 
\multicolumn{1}{c|}{\textcolor{blue}{2}} & 
\multicolumn{1}{c|}{\textcolor{OliveGreen}{3}} &
\multicolumn{1}{c|}{4} \\ \hline

\multicolumn{1}{|l|}{$n_0$ [$10^{18}$ cm$^{-3}$] } & 1.1 & 0.9 & 1.1 & 0.8\\ 
\hline
\multicolumn{1}{|l|}{$a_0$ } & 2.96 & 2.64 & 3.24 & 2.96\\ 
\hline

\multicolumn{1}{|l|}{$P/P_{cr}$ } & 18.3 & 12.0 & 22.0 & 13.3\\ 
\hline
\multicolumn{1}{|l|}{Laser Energy [J]} & 10 & 10 & 15 & 15\\ 
\hline
\multicolumn{1}{|l|}{Pulse duration [fs] } & 20 & 25 & 25 & 30\\ 
\hline
\multicolumn{1}{|l|}{Laser propagation distance [mm]} & 6.3 & 7.2 & 6.5 & 7.6 \\ 
\hline
\multicolumn{1}{|l|}{$Q_{beam}$ [pC]} & 263 & 48 & 543 & 24\\ 
\hline
\multicolumn{1}{|l|}{$Q_{total,\thinspace >300\thinspace MeV}$ [pC]} & 426 & 72 & 729 & 24\\ 
\hline
\multicolumn{1}{|l|}{$E_{spectrum\thinspace peak}$ [MeV]} & 870 & 740 & 930 & 1130\\ 
\hline
\multicolumn{1}{|l|}{$\Delta E/E_{spectrum\thinspace peak}$ [\%]} & 8.3 & 3.2 & 6.4  & 2.0\\ 
\hline
\multicolumn{1}{|l|}{$2\sigma_{y}$ [$\mu$m]} & 3.0 & 2.3 & 2.9 & 0.5\\ 
\hline
\multicolumn{1}{|l|}{$2\sigma_{z}$ [$\mu$m]} & 2.8 & 2.3 & 3.1 & 0.5\\ 
\hline
\multicolumn{1}{|l|}{$\varepsilon_{n,y}$ [mm-mrad]} & 14.5 & 1.3 & 12.3 & 0.4\\ 
\hline
\multicolumn{1}{|l|}{$\varepsilon_{n,z}$ [mm-mrad]} & 11.8 & 1.3 & 13.1 & 0.4\\ 
\hline

\end{tabular}
\caption{Resume of the plasma plateau density $n_0$, initial laser pulse parameters used in the set of first stage simulations. 
For the electron beam, the peak in the energy spectrum $E_{spectrum\thinspace peak}$, the relative energy spread $\Delta E/E_{spectrum\thinspace peak}$ , rms size (2 standard deviations $\sigma$) and the normalized emittances $\varepsilon_{n,i}$ ($i=y,\thinspace z$) in the transverse planes are reported. These parameters are measured after a certain laser propagation distance, before the phase space rotation, reported in the table.}
\label{table:beam_params_1st_stage}
\end{table}

Even in the relatively well constrained range of parameters, significant differences are observed.
First, there is a factor 20 between the minimum and maximum beam charges.
This parameter appears to be controlled primarily by the $P/P_{\rm cr}$ ratio.
Lasers close to the $P/P_{\rm cr} = 10$ are subject to milder self-focusing and trigger  less self-injection (simulations 2 and 4).
Conversely, for higher  $P/P_{\rm cr}$, self-injection is dramatically enhanced (simulations 1 and 3).
This dependency is expected \cite{beck2014} and provides a practical way to optically control the injected charge even though the sensitivity is such that it might be difficult to achieve a very good accuracy.

The energy evolution is roughly similar in all cases with the exception of simulation 2.
This evolution is governed by the accelerating field $E_{\rm acc}$.
In the bubble regime, $E_{\rm acc}$ is proportional to $\sqrt{a_0n_0}$\cite{Esarey2009}.
In the simulations, $n_0$ is constant and therefore, $a_0$ evolution dictates the evolution of the accelerating field.
This evolution is shown as the normalized peak $|\tilde{E}|$ in figure \ref{fig:a0_E_Q}.
Simulation 2 has a small density and the weakest $a_0$ throughout the run because of a low inital $a_0$ but also the weakest self-focusing.
This explains the small acceleration observed in that case with respect to the other simulations.
Simulation 4, performs quite well in terms of acceleration in spite of having the smallest density and only a medium initial $a_0$ thanks to a superior laser guiding. 
Simulation 3, in spite of having the highest $a_0$ and density do not outperform other simulations in terms of energy because in that case the beam loading effect is not negligible any more\cite{Tzoufras2008}.

The computing time needed by the 3D simulations for 1 mm is of the order of 30-60 kh. 
We remark that a parameter scan of four 12 mm long 3D simulations as the one presented in this work would have been extremely costly with standard laser simulations, which are typically slower by at least a factor 20. This justifies the interest in time-explicit envelope models \cite{Benedetti2012, Cowan2011}, especially for preliminary studies for experiments like those planned for Apollon.


\section{Conclusions}
A recently published, time explicit, easily parallelizable envelope method for PIC codes has the potential to model laser-plasma interaction involved in many LWFA setups with considerable speedups compared to standard laser simulations. The implementation of this time-explicit model in \Smilei has been described. The suitability of this envelope model for single-stage with self-injection and second-stage LWFA simulations has been explored. The laser and plasma parameters chosen for these studies lie within the intervals of interest for the upcoming Apollon LWFA experiments.

The envelope model used in this work has been benchmarked against a standard laser simulation in a scenario with external injection of a witness electron beam, yielding a speedup of 20 compared to a standard laser simulation and a good agreement even at a distance of 15 mm. Besides, the envelope model allows to model more accurately the longitudinal and transverse phase space evolution of the injected beam. Indeed, in the longitudinal direction a more accurate prediction of the laser propagation speed by the envelope equation reduces the numerical dephasing of the electron beam caused by FDTD solvers of Maxwell's equations. In the transverse direction, the envelope simulations allow for low-pass filtering which in this case considerably reduced the growth of NCR, conserving the beam emittance.

In the single-stage simulations, four potential working points in the regions of parameters of interest have been tested, finding the injection of $\approx1$ GeV electron beams with energy spread lower than $10\%$. 
These working points have been found examining the evolution of the injection and acceleration process for 12 mm in four 3D simulation, a propagation distance which would need considerable computing resources to be simulated with a standard 3D PIC code.

The use of this envelope model may thus represent an important investigation tool for the study of LWFA and parameter explorations with reduced resource requirements. 
Further improvements could include spectral solvers for the envelope equation and Maxwell's equations to cope with numerical Cherenkov radiation and the use of cylindrical symmetry to speed up even further the simulations (albeit losing the full 3D characteristics of the phenomena).

\appendix
\section{Envelope PIC loop}\label{appendix}
We report a brief summary of the numerical solution of the equations involved in the envelope model used for the simulations of this work. More details on the derivation of the equations and the numerical schemes used to solve them can be found in \cite{Terzani2019}.
The envelope equation that \Smilei solves is
\begin{equation}\label{envelope_equation2}
\nabla^2 \tilde{A}+2i\left(\partial_x \tilde{A} + \partial_t \tilde{A}\right)-\partial^2_t\tilde{A}=\chi \tilde{A},\quad
\end{equation}
in which the the susceptibility $\chi$ is defined as
\begin{equation}\label{susceptibility}
  \chi(\mathbf{x}) = \sum_s\,r_s^2\,\sum_p\,\frac{w_p}{\bar{\gamma}_p}\,S\big(\mathbf{x}-\mathbf{\bar{x}}_p\big)\ ,
\end{equation}
where $\bar{\gamma}_p$ is the averaged or ponderomotive Lorentz factor of the macroparticle $p$ and $r_s=q_s/m_s$ is the charge to mass ratio ($q_s$ and $m_s$ are respectively the particle species $s$ charge and mass normalized by the elementary charge $e$ and electron mass $m_e$). The MP weight and shape factor are denoted by $w_p$ and $S(\mathbf{x})$ respectively (see \cite{BirdsallLangdon2004} for the definition of shape factor). The averaged Lorentz factor is defined from the averaged particle momentum $\mathbf{\bar{u}}_p=\mathbf{\bar{p}}_p/m_s$ and the ponderomotive potential $\Phi=|\tilde{A}|^2/2$ \cite{Quesnel1998,Mora1997,Cowan2011}:

\begin{equation} \label{gamma_ponderomotive}
\bar{\gamma}_p = \sqrt{1+\mathbf{\bar{u}}_p\cdot\mathbf{\bar{u}}_p+r_s^2\Phi(\mathbf{\bar{x}}_p)}.
\end{equation}

Maxwell's equations retain their form, except for substituting the electromagnetic fields and the source terms with their respective low frequency components (denoted with a bar):
\begin{eqnarray}\label{Maxwell_envelope}
\partial_t \mathbf{\bar{E}} &=& \nabla \times \mathbf{\bar{B}} - \mathbf{\bar{J}}  \\
\partial_t \mathbf{\bar{B}} &=& -\nabla \times \mathbf{\bar{E}}   .
\end{eqnarray}
The low frequency current density $\mathbf{\bar{J}}$ is projected on the grid through Esirkepov's method \cite{esirkepov:CPC2001}.

Also the MP equations of motion remain similar to the usual ones, but with a crucial difference: the low frequency components of the position and momentum are pushed by both the low frequency electromagnetic fields (the Lorentz force term) and the ponderomotive force $\mathbf{F}_{pond}=-r^2_s\thinspace\frac{1}{4\bar{\gamma}_p}\nabla\Phi$. This term takes into account the averaged effect of the laser on the particles and can be computed from the envelope $\tilde{A}$. The resulting equations of motions for the particle of index $p$ are:

\begin{eqnarray}\label{ponderomotive_equations_of_motion}
\frac{d\mathbf{\bar{x}}_p}{dt} &=& \frac{\mathbf{\bar{u}}_p}{\bar{\gamma}_p}\,\\
\frac{d\mathbf{\bar{u}}_p}{dt} &=& r_s \, \left( \mathbf{\bar{E}}_p + \frac{\mathbf{\bar{u}}_p}{\bar{\gamma}_p} \times \mathbf{\bar{B}}_p \right)-r^2_s\thinspace\frac{1}{4\bar{\gamma}_p}\nabla\Phi_p.
\end{eqnarray}

\begin{figure}[htbp]
\begin{center}
\includegraphics[scale=0.6]{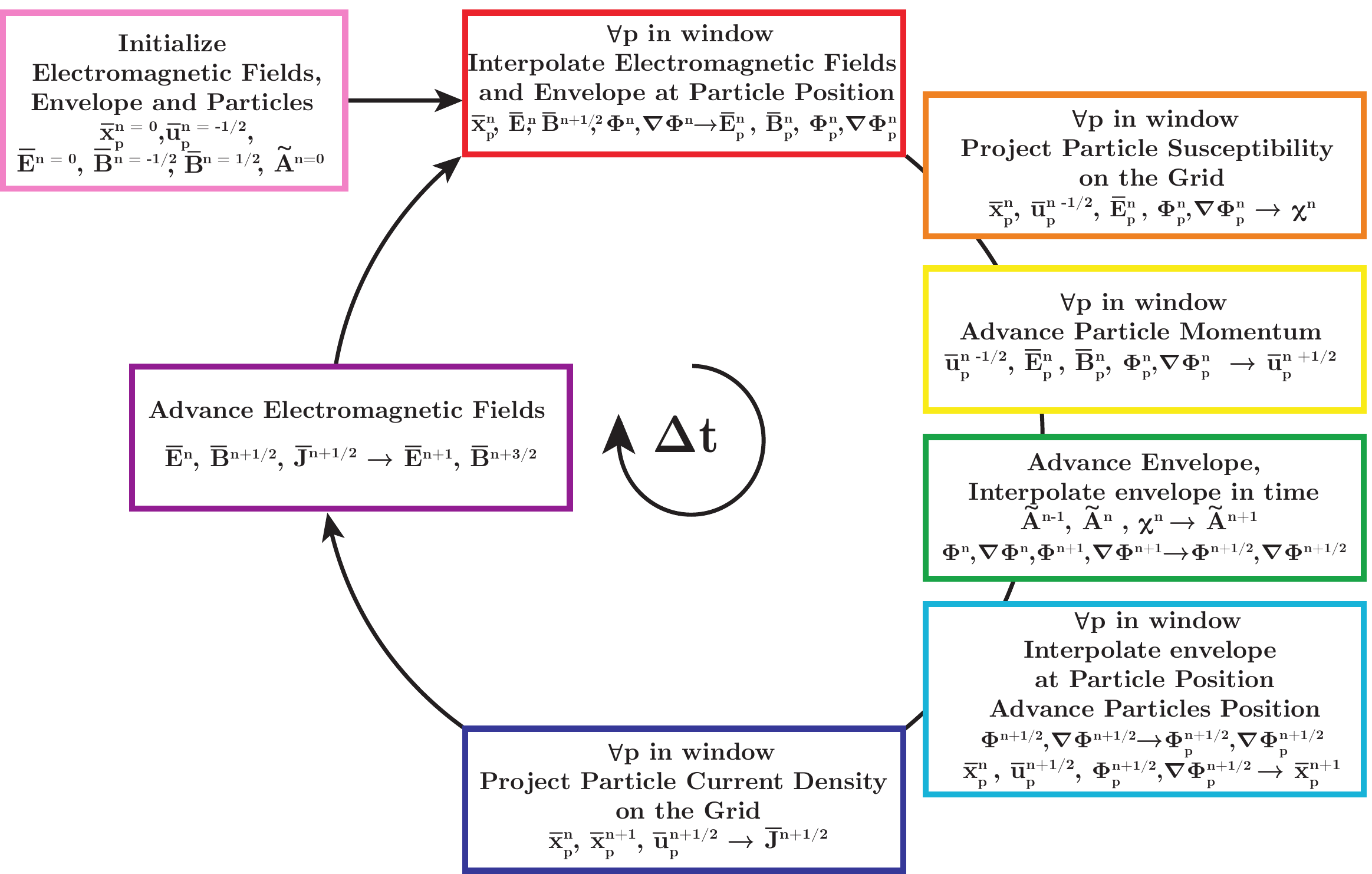}
\caption{Ponderomotive Particle in Cell loop. The definition of ponderomotive potential $\Phi=\frac{|\tilde{A}|^2}{2}$ is used.}
\label{fig:Envelope_Loop}
\end{center}
\end{figure}

The nonlinearities introduced in the envelope source term (Eqs. \ref{susceptibility},\ref{gamma_ponderomotive} ) and the ponderomotive force by the terms $\bar{\gamma}$ (Eq. \ref{gamma_ponderomotive}) and $\nabla\Phi$ require a modification of the standard PIC algorithm.
The PIC loop is changed to implement the solution of the ponderomotive equations (Eqs. \ref{envelope_equation2},\ref{ponderomotive_equations_of_motion}), adding the envelope equation solver, the susceptibility deposition and a ponderomotive particle push, as depicted in Fig.\ref{fig:Envelope_Loop}.

The interpolation of the fields to the MP positions and the deposition of the current density and susceptibility on the grid are implemented as in standard PIC codes \cite{BirdsallLangdon2004}. In the simulations presented in this work, we used an order 2 shape function, without loss of generality.

As explained in detail in \cite{Terzani2019}, Eq. \ref{envelope_equation} can be discretized using finite differences both in time and space. The second order centered finite differences yield an explicit scheme:

\begin{eqnarray}\label{envelope_solver}
\tilde{A}_{ijk}^{n+1}= \frac{1+i\Delta t}{1+\Delta t^2} \Bigg[2 \tilde{A}_{ijk}^n-(1+i\Delta t)\tilde{A}_{ijk}^{n-1}+\Bigg(\nabla^2\tilde{A}_{ijk}\bigg|_{ijk}^n-\chi^n_{ijk}\tilde{A}_{ijk}^n+2i\Delta t^2 \frac{\tilde{A}_{i+1\thinspace jk}^{n}-\tilde{A}_{i-1\thinspace jk}^{n}}{2\Delta z}\Bigg) \Bigg] \quad
\end{eqnarray}

where
\begin{eqnarray}
\nabla^2\tilde{A}\bigg|_{ijk}^n=&\frac{\tilde{A}_{i+1 \thinspace jk}^{n}-2\tilde{A}_{ijk}^{n}+\tilde{A}_{i-1\thinspace jk}^{n}}{\Delta x^2}+\frac{\tilde{A}_{ij-1\thinspace k}^{n}-2\tilde{A}_{ijk}^{n}+\tilde{A}_{ij+1\thinspace k}^{n}}{\Delta y^2}+\frac{\tilde{A}_{ijk-1}^{n}-2\tilde{A}_{ijk}^{n}+\tilde{A}_{ijk+1}^{n}}{\Delta z^2}.
\end{eqnarray}
The indices $i$,$j$,$k$ refer to the mesh cell indices along the $x$, $y$, $z$ directions. The envelope, the ponderomotive potential and the susceptibility are centered in space as the charge density in all directions and time-centered as the electric field in the Yee scheme \cite{Yee1966} .

Once the susceptibility term at time step $n$ is known, the past and present envelope fields ($\tilde{A}^{n-1}$, $\tilde{A}^n$ respectively) can be used to compute $\tilde{A}^{n+1}$ using Eq. \ref{envelope_solver}. The stencil of this explicit envelope solver contains three points in space for each direction.

To project the susceptibility on the grid and to update the MP momenta , as explained in \cite{Terzani2019}, the ponderomotive Lorentz factor at the timestep $n$ is necessary. This quantity cannot be directly computed from Eq. \ref{gamma_ponderomotive}, since the MP momentum is known only at the timestep $n-1/2$. Thus, a two step computation (derived in \cite{Terzani2019}) is used to obtain an approximation of the desired Lorentz factor $\bar{\gamma}_p^n$:
\begin{eqnarray}
\gamma_{0p}^2 &=& 1+r_s^2\Phi_p^n +\mathbf{\bar{u}}_p^{n-1/2}\cdot\mathbf{\bar{u}}_p^{n-1/2},\label{gamma_0}\\
\bar{\gamma}_p^n &=&\gamma_{0p}+\frac{1}{2\gamma_{0p}^2}\Bigg( \gamma_{0p} \frac{\Delta t}{2} r_s \mathbf{\bar{E}}^n - \frac{\Delta t}{4} r_s^2 \nabla\Phi_p^n \Bigg)\cdot\mathbf{\bar{u}}_p^{n-1/2}.\label{lorentz_factor_pusher}
\end{eqnarray}
This ponderomotive Lorentz factor $\bar{\gamma}_p^n$ is used in the deposition of susceptibility, following Eq. \ref{susceptibility}.

A modified Boris pusher \cite{BirdsallLangdon2004} can be used to update the MP momenta. The only modifications to the standard scheme are the use of $\left( r_s\mathbf{\bar{E}}_p^n-r_s^2\frac{\nabla\Phi^n_p }{2\bar{\gamma}_p^n}      \right)$ instead of $r_s\mathbf{\bar{E}}_p^n$ as the source term changing the particle $p$ energy and the term $r_s\frac{\mathbf{\bar{B} }^n_p}{\bar{\gamma}_p^n}$ for its momentum rotation under the effect of the magnetic field. The ponderomotive Lorentz factor $\bar{\gamma}_p^n$ used by the pusher is the one from Eq. \ref{lorentz_factor_pusher}.

Again, to update the particle $p$ position through a leapfrog scheme, the necessary ponderomotive Lorentz factor at timestep $n+1/2$ cannot be computed directly from Eq. \ref{gamma_ponderomotive}, since the ponderomotive potential $\Phi$ and its gradient can be interpolated only at the known MP position at timestep $n$. Hence, a two-step approximation of the ponderomotive Lorentz factor $\bar{\gamma}_p^{n+1/2}$ (derived in \cite{Terzani2019}) is used:
\begin{eqnarray}
\gamma_{0p}^2 &=& 1+\mathbf{\bar{u}}_p^{n+1/2}\cdot \mathbf{\bar{u}}_p^{n+1/2}+ r_s^2\Phi_p^{n+1/2}(\mathbf{\bar{x}}_p^{n}),\label{gamma1}\\
\bar{\gamma}^{n+1/2}_p &=& \gamma_{0p}+\frac{\Delta t}{4\gamma^2_{0p}} \mathbf{\bar{u}}_p^{n+1/2}\cdot r_s^2\nabla\Phi_p^{n+1/2}(\mathbf{\bar{x}}_p^{n}).\label{gamma2}
\end{eqnarray}

Note that the ponderomotive potential and its gradient in Eqs. \ref{gamma1}, \ref{gamma2} are defined at the timestep $n+1/2$. Since these quantities known at the timesteps $n$ and $n+1$, their value at $n+1/2$ can be obtained through linear interpolation. After this interpolation in time, they are interpolated at the known particle position at the timestep $n$ and then used to compute $\bar{\gamma}_p^{n+1/2}$ through Eqs. \ref{gamma1}, \ref{gamma2}.

Finally, using $\bar{\gamma}_p^{n+1/2}$ of Eq. \ref{gamma2}, the updated position can be computed:
\begin{equation}
\mathbf{\bar{x}}_p^{n+1}=\mathbf{\bar{x}}_p^{n}+\frac{\mathbf{\bar{u}}_p^{n+1/2}}{\bar{\gamma}^{n+1/2}_p}\Delta t.
\end{equation}

\section*{Acknowledgments}
F. Massimo was supported by P2IO LabEx (ANR-10-LABX-0038) in the framework “Investissements d’Avenir” (ANR-11-IDEX-0003-01) managed by the Agence Nationale de la Recherche (ANR, France).
This work was granted access to the HPC resources of TGCC/CINES under the allocation 2018-A0050510062 and Grand Challenge “Irene" 2018 project gch0313 made by GENCI. The authors are grateful to the TGCC and CINES engineers for their support. The authors thank the engineers of the LLR HPC clusters for resources and help. The authors are grateful to the ALaDyn development team for the help and discussions during the development of the envelope model, in particular D. Terzani, A. Marocchino and S. Sinigardi and to Gilles Maynard for fruitful discussions.

%
%
%
%
%

\renewcommand{\baselinestretch}{1.0}

\section*{Bibliography}
\bibliographystyle{unsrt}
\bibliography{Bibliography}

\end{document}